\documentclass[twocolumn]{aa}
\usepackage{graphicx}
\usepackage{natbib}
\bibpunct{(}{)}{;}{a}{}{,}

\newcommand{\id}{\mbox{$\mathrm{d^{-1}}$}}
\newcommand{\he}[1] {He\,{\sc #1}}

\newcommand{\kms}{\mbox{$\mathrm{km~s^{-1}}$}}
\newcommand{\Porb}{\mbox{$P_\mathrm{orb}$}}
\newcommand{\Pspin}{\mbox{$P_\mathrm{spin}$}}
\newcommand{\Line}[3]{\Ion{#1}{#2}~$\lambda$#3}

\newcommand{\Ion}[2]{#1{\,\scriptsize #2}}

\newcommand{\Ha}{\mbox{${\mathrm H\alpha}$}}
\newcommand{\Hb}{\mbox{${\mathrm H\beta}$}}
\newcommand{\Hg}{\mbox{${\mathrm H\gamma}$}}
\newcommand{\Hd}{\mbox{${\mathrm H\delta}$}}
\newcommand{\Rwd}{\mbox{$R_{\mathrm{wd}}$}}
\newcommand{\Msun}{\mbox{$M_{\odot}$}}
\newcommand{\Mwd}{\mbox{$M_{\mathrm{wd}}$}}

\begin{document}
\title{\object{HS\,0943+1404}, a true intermediate polar\thanks{Based
in part on observations obtained at the German-Spanish Astronomical
Center, Calar Alto, operated by the Max-Planck-Institut f\"{u}r
Astronomie, Heidelberg, jointly with the Spanish National Commission
for Astronomy, on observations made with the NOT telescope, operated
on the island of La Palma by the Instituto de Astrof\'\i sica de
Canarias (IAC) at the Spanish Observatorio del Roque de los Muchachos,
on observations made with the IAC80 telescope, operated on the island
of Tenerife by the IAC at the Spanish Observatorio del Teide, on
observations made with the 1.2\,m telescope at the FLWO Observatory, a facility of the Smithsonian Institution.}}

\author{P. Rodr\'\i guez-Gil\inst{1,2} \and
        B. T. G\"ansicke\inst{1}\and
	H.-J. Hagen\inst{3}\and
	D. Nogami\inst{4}\and
	M. A. P. Torres\inst{5}\and
	H. Lehto\inst{6,7}\and
	A. Aungwerojwit\inst{1}\and
	S. Littlefair\inst{8}\and
	S. Araujo-Betancor\inst{9}\and
	D. Engels\inst{3}}

\offprints{P. Rodr\'\i guez-Gil,\\\email{prguez@iac.es}}

\institute{
  Department of Physics, University of Warwick, Coventry CV4 7AL, UK
  \and
  Instituto de Astrof\'\i sica de Canarias, V\'\i a L\'actea, s/n, La Laguna, E-38205 Santa Cruz de Tenerife, Spain
\and
  Hamburger Sternwarte, Universit\"at Hamburg, Gojenbergsweg 112,
  21029 Hamburg, Germany
\and
  Hida Observatory, Kyoto University, Kamitakara, Gifu 506-1314, Japan
\and
Harvard-Smithsonian Center for Astrophysics, 60 Garden St, Cambridge, MA 02138, USA
\and
Tuorla Observatory, Turku University, V\"ais\"al\"antie 20, 21500 Piikki\"o, Finland
\and
Department of Physics, 20014 University of Turku, Finland
\and  
School of Physics, University of Exeter, Exeter, EX4 4QL, UK
\and
Space Telescope Science Institute, 3700 San Martin Drive, Baltimore, MD 21218, USA
 }

\date{Received 2005; accepted 2005}

\abstract{We have identified a new intermediate polar, HS\,0943+1404,
  as part of our ongoing search for cataclysmic variables in the
  Hamburg Quasar Survey. The orbital and white dwarf spin periods
  determined from time-resolved photometry and spectroscopy are
  $P_\mathrm{orb} \simeq 250$\,min and  $\Pspin = 69.171 \pm 0.001$\,min,
  respectively. The combination of a large ratio $\Pspin/\Porb\simeq0.3$
  and a long orbital period is very unusual compared to the other
  known intermediate polars. The magnetic moment of the white dwarf is
  estimated to be $\mu_1\sim10^{34}\,\mathrm{G\,cm^{3}}$, which is in
  the typical range of polars. Our extensive photometry shows that
  HS\,0943+1404 enters into deep ($\sim3$\,mag) low states, which are
  also a characteristic feature of polars. We therefore suggest that the
  system is a true ``intermediate'' polar that will eventually
  synchronise, that is, a transitional object between intermediate polars and polars. The optical spectrum of HS\,0943+1404 also exhibits a number
  of unusual emission lines, most noticeably \Line{N}{II}{5680}, which
  is likely to reflect enhanced nitrogen abundances in the envelope of
  the secondary. 
 \keywords{accretion, accretion discs -- binaries: close --
stars: individual: HS\,0943+1404 -- novae, cataclysmic variables}}

\titlerunning{The new intermediate polar HS\,0943+1404}
\authorrunning{P. Rodr\'\i guez-Gil et al.}

\maketitle

\section{Introduction}

Our understanding of the evolution of cataclysmic variables (CVs) is
still very fragmentary. It is clear that the current theory of CV
evolution~---~largely resting on the disrupted magnetic braking
scenario \citep{king88-1}~---~makes a number of predictions in strong
conflict with the observational data. For instance, the theoretical minimum orbital period is $\sim 70$ min, much shorter than the observed $\sim 80$ min \citep{kolb+baraffe99-1}. In addition, CVs are expected to spend most of their lifetime near the minimum period, a fact not corroborated by the observations. With regard to the CV populations above and below the period gap, we observe similar numbers of systems on both sides, while all population syntheses predict that $\sim 95$~\% of the entire CV population should have orbital periods below the gap \citep{kolb93-1,howelletal97-1}. These three fundamental discrepancies, among others, clearly depict the current misaligning between theory and observations. 

Although CV evolution theory has
undergone a number of modifications/alternatives
\citep[e.g.][]{king+schenker02-1, schenker+king02-1, schenkeretal02-1,
andronovetal03-1}, the discrepancies with the properties of the
observed CV population remain unsatisfactorily large, and it seems to
be clear that observational selections effects represent a
major problem. \citet{gaensicke04-1} has shown that the
majority of all CVs have been identified primarily through their
variability or through the detection of X-rays, leaving a large
parameter space unsampled. We are currently carrying out a large-scale search
for CVs based on their spectroscopic properties in the Hamburg Quasar
Survey (HQS, \citealt{hagenetal95-1}), with the aim of identifying CVs
that are inconspicuous in variability or X-ray surveys
\citep{gaensickeetal02-2}.

This method proved successful and led to the discovery of a number of
systems that show either no or infrequent outbursts, such as e.g. the
dwarf nova GY\,Cnc\,=\,HS\,0907+1902 \citep{gaensickeetal00-2}, the
SW\,Sex star HS\,0728+6738 \citep{rodriguez-giletal04-2}, and the
enigmatic low-mass transfer system HS\,2331+3905
\citep{araujo-betancoretal05-1}. Another important class of CVs
characterised by rather low levels of variability are those systems
containing a magnetic white dwarf, that is rotating either synchronous
to the orbital period (polars) or asynchronous (intermediate polars,
IPs). The follow-up of HQS CV candidates led to the discovery of a
new type of very low mass transfer polars, WX\,LMi\,=\,HS\,1023+3900
\citep{reimersetal99-1} and HS\,0922+1333 \citep{reimers+hagen00-1},
that may be systems shortly before the onset of Roche-lobe overflow
from the secondary; the ultra high-field polar
RX\,J1554.2+2721\,=\,HS\,1552+2730 \citep{jiangetal00-1,
thorstensen+fenton02-1, gaensickeetal04-1}; and the two intermediate
polars 1RXS\,J062518.2+733433\,=\,HS\,0618+7336
\citep{araujo-betancoretal03-2} and DW\,Cnc\,=\,HS\,0756+1624
\citep{rodriguez-giletal04-1}.

Here we report on the discovery of the third IP, 
HS\,0943+1404, which exhibits very unusual characteristics, and may be
one example of the long-sought link between asynchronous and
synchronous magnetic CVs. 

\section{Observations and data reduction}

\begin{figure}
\includegraphics[width=\columnwidth]{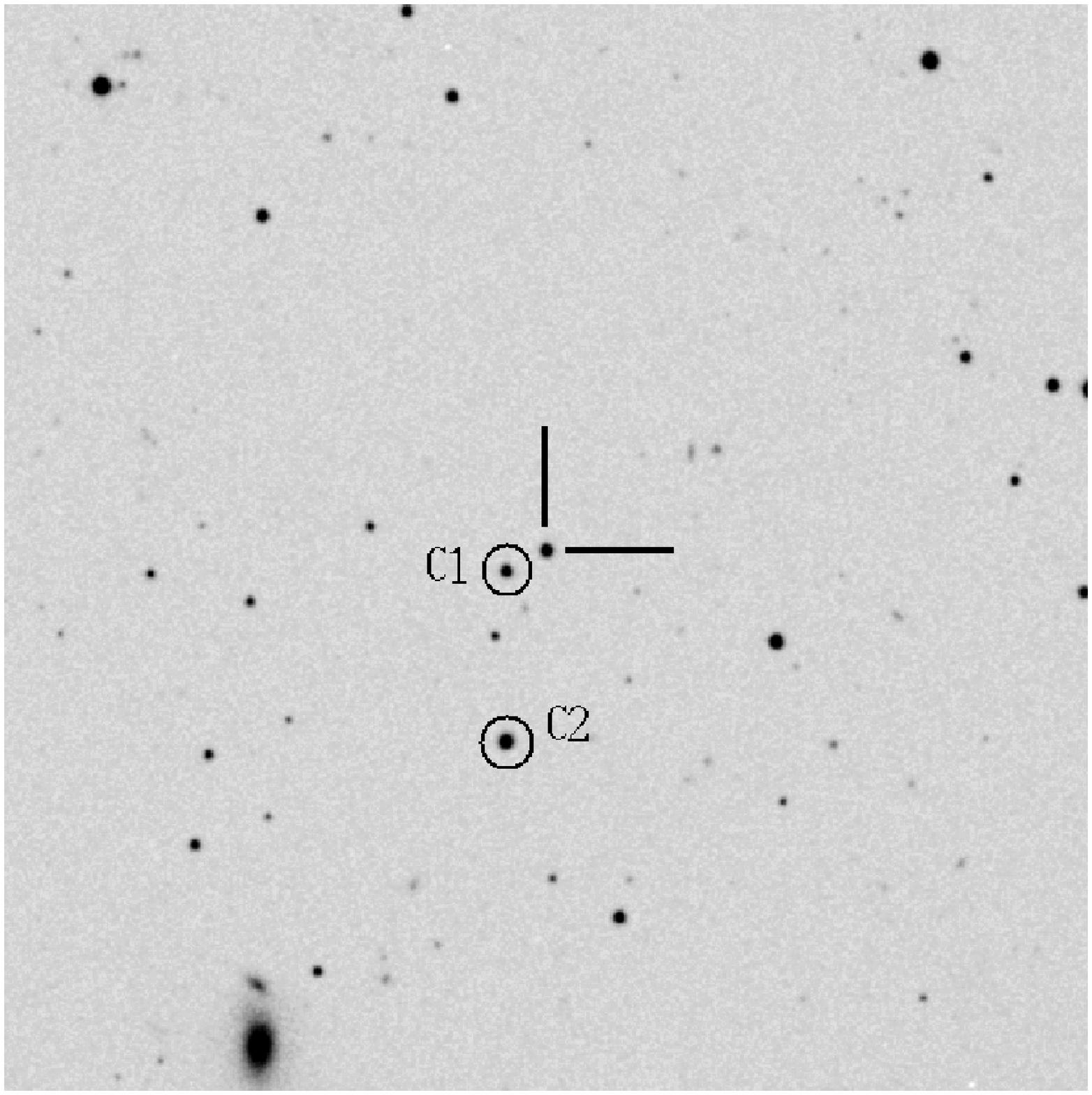}
\caption[]{\label{f-fc} $10\arcmin\times10\arcmin$ finding chart of
HS\,0943+1404 obtained from the Digitized Sky Survey. The
coordinates of the CV are
$\alpha(\mathrm{J}2000)=09^\mathrm{h}46^\mathrm{m}34.5^\mathrm{s}$,
$\delta(\mathrm{J}2000)=+13\degr50\arcmin58.1\arcsec$. North is up and east to the left. `C1' is the primary comparison star and `C2' is the check star.}
\end{figure}

\begin{table}
\setlength{\tabcolsep}{1.1ex}
\caption[]{\label{t-obslogphot}Log of the photometric observations}
\begin{flushleft}
\begin{tabular}{lcccccc}
\hline\noalign{\smallskip}
UT Date & Coverage & Filter & Exp. & \#\,Frames & Mean\\
         &   (h)   &        &   (s)&            & magnitude\\  
\hline\noalign{\smallskip}
\multicolumn{5}{l}{\textbf{Calar Alto 2.2\,m}} \\
2001 Apr 30    &    - &  $V$    & 30 &  1  & 17.0 \\
2002 Dec 14    &    - &  $V$    & 30 &  1  & 17.6 \\
2002 Dec 15    &    - &  $V$    & 30 &  1  & 17.6 \\
2002 Dec 28    & 1.16 &  $V$    & 45 & 64  & 19.2 \\
2003 Mar 06    &    - &  Clear  & 20 &  1  & 19.1 \\
2003 Dec 17    & 1.11 &  Clear  & 30 & 88  & 16.1 \\
2003 Dec 23    &    - &  Clear  & 30 &  2  & 16.6 \\
2003 Dec 23    &    - &  $V$    & 30 &  2  & 16.5 \\
\noalign{\smallskip}
\multicolumn{5}{l}{\textbf{Calar Alto 3.5\,m,}} \\
2004 Mar 02    &    - & $V$     & 30 &  1  & 16.7 \\
2004 Mar 03    &    - & $V$     & 30 &  1  & 16.5 \\
\noalign{\smallskip}
\multicolumn{5}{l}{\textbf{FLWO 1.2\,m, CCD}} \\
2004 Mar 09  & 4.96 &  Clear  & 60 & 407 & 16.2  \\
2004 Mar 10  & 7.70 &  Clear  & 60 & 471 & 16.3  \\
2004 Mar 19  & 3.72 &  Clear  & 60 & 270 & 16.4  \\
2004 Mar 20  & 7.08 &  Clear  & 60 & 605 & 16.3  \\
\noalign{\smallskip}
\multicolumn{5}{l}{\textbf{IAC80 0.8\,m, CCD}} \\
2003 Dec 27    & 5.81 &  Clear  & 40 & 500 & 16.3 \\
2003 Dec 28    & 3.19 &  Clear  & 40 & 261 & 16.4 \\
\noalign{\smallskip}
\multicolumn{5}{l}{\textbf{Tuorla 0.7\,m, CCD}} \\
2004 Jan 20  & 3.97 &  Clear  & 60 & 217 & 16.5 \\
2004 Feb 09  & 6.89 &  Clear  & 60 & 360 & 16.4 \\
2004 Mar 16  & 5.76 &  Clear  & 60 & 299 & 16.5 \\
2004 Mar 18  & 7.83 &  Clear  & 60 & 395 & 16.5 \\
\noalign{\smallskip}
\multicolumn{5}{l}{\textbf{Hida 0.6\,m, CCD}} \\
2004 Mar 09  & 6.64 &  Clear  & 60 & 321 & 16.4 \\
2004 Mar 13  & 2.26 &  Clear  & 60 & 113 & 16.3 \\
2004 Mar 15  & 3.67 &  Clear  & 60 & 191 & 16.4 \\
2004 Mar 19  & 2.67 &  Clear  & 60 & 136 & 16.4 \\
2004 Mar 26  & 0.52 &  Clear  & 60 & 24  & 16.5 \\
2004 Mar 27  & 6.17 &  Clear  & 60 & 296 & 16.3 \\
2004 Mar 28  & 6.60 &  Clear  & 60 & 306 & 16.4 \\
\noalign{\smallskip}
\multicolumn{5}{l}{\textbf{JKT 1.0\,m}} \\
2002 Dec 29  & 2.72 & Clear & 60 & 120 & 19.0 \\
2002 Dec 31  & 6.25 & Clear & 60 & 296 & 19.0 \\
2003 Jan 04  & 2.75 & Clear & 45 & 148 & 18.2 \\
2003 Jan 05  & 0.40 & Clear & 30 &  32 & 18.7 \\
2003 Jan 07  & 3.52 & Clear & 45 & 205 & 18.9 \\
2003 Jan 09  & 2.76 & Clear & 60 &  91 & 19.1 \\
\noalign{\smallskip}\hline
\end{tabular}
\end{flushleft}
\end{table}

\subsection{Photometry}
Differential CCD photometric time series of HS\,0943+1404
were obtained during 25 nights throughout the period December 2002 to
March 2004, accumulating a total coverage of $\simeq106$\,h
(Table~\ref{t-obslogphot}). The photometric observations were carried
out at six different telescopes. Brief details about the individual
instrumentation and reduction process are given below.

\paragraph{Calar Alto Observatory.}
At Calar Alto, filterless and $V$-band CCD photometry of HS\,0943+1404
was obtained in December 2002 and December 2003 with the CAFOS SITe
CCD camera on the 2.2\,m telescope using a small read-out window. The
photometric reduction was carried out using the pipeline described by
\citet{gaensickeetal04-1}, which, in brief, pre-processes the images
in \texttt{ESO-MIDAS} and performs aperture photometry of all stars in
the field of view using the \texttt{Sextractor}
\citep{bertin+arnouts96-1}. Differential magnitudes of HS\,0943+1404
were measured relative to the star `C1' (USNO--A2.0~0975--06331245,
$R=15.9$, $B=15.9$; see Fig.\,\ref{f-fc}), located
$\simeq30\arcsec$ south-east of the CV. The comparison star `C1' was checked against the
secondary comparison star `C2' (USNO--A2.0~0975--06331238, $R=14.3$,
$B=14.6$), and no significant variability was detected. In addition, a
number of acquisition images of HS\,0943+1404 were obtained prior to
spectroscopic observations (Sect.\,\ref{s-spect}), at both the 2.2\,m
and the 3.5\,m telescopes, and the magnitude of the target was
determined interactively with the \texttt{IRAF} task \texttt{imexam},
using again the comparison star `C1' in Fig.\,\ref{f-fc}.

\paragraph{Observatorio del Roque de los Muchachos.}
On La Palma, filterless CCD photometry of HS\,0943+1404 was obtained
in December 2002/January 2003 using the Jacobus Kapteyn Telescope
(JKT) with the $2\,\mathrm{k}\times2\,\mathrm{k}$ pixel$^2$ SITe
detector. Observations and data reduction were carried out in an
analogous fashion as described above for the Calar Alto photometry.

\paragraph{Fred Lawrence Whipple Observatory.}
Filterless observations of HS\,0943+1404 were carried out with the
1.2\,m telescope at the Fred Lawrence Whipple Observatory (FLWO) in
March 2004. The 4-Shooter CCD camera was in place, which consists of
an array of four 2048$\times$2048 pixel$^2$ CCDs. Only a small window of the CCD \#3 was read out. Data reduction was carried out using
\texttt{IRAF}. After bias and flat-field
corrections, the images were aligned and instrumental magnitudes of
HS\,0943+1404 and the comparison stars `C1' and `C2' were extracted
using the point spread function (PSF) packages. Differential
magnitudes of HS\,0943+1404 were then computed relative to `C1'.

\paragraph{Observatorio del Teide.} Filterless CCD photometry of
HS\,0943+1404 was obtained at the 0.8\,m IAC80 telescope at the
Observatorio del Teide on Tenerife in December 2003, using the Thomson
$1\,\mathrm{k}\times1\,\mathrm{k}$ pixel$^2$ camera. The observations
and data reduction were carried out in an analogous way as described
above for the FLWO photometry.

\paragraph{Tuorla Observatory.}
The 0.7\,m Schmidt-Vaisala telescope at Tuorla Observatory was used in
January, February and March 2004 to obtain filterless photometry of
HS\,0943+1404 using a SBIG ST--8 CCD camera. Observations and data
reduction were carried out in the same way as described above for the
Calar Alto photometry.

\paragraph{Hida Observatory}
We performed filterless CCD photometry of HS\,0943+1404 with
the 0.6\,m telescope and the SITe CCD camera at Hida Observatory (Japan) in March 2004. After standard reduction of the raw images, differential aperture photometry was carried out relative to the comparison star `C1' in a standard manner using \texttt{IRAF}.

\begin{table}[t]
\setlength{\tabcolsep}{1.1ex}
\caption{\label{t-obslogspec}Log of the spectroscopic observations}
\begin{flushleft}
\begin{tabular}{lcccc}
\hline\noalign{\smallskip}
UT Date & Coverage & Grating & Exp. & \#\,Frames \\
         &   (h)   &                &   (s) & \\  
\hline\noalign{\smallskip}
\multicolumn{5}{l}{\textbf{Calar Alto 2.2\,m, CAFOS}} \\
2001 Apr 30    &    -   & B--200 & 600 & 1 \\
2002 Dec 14    & 0.75   & G--100 & 600 & 5 \\
2002 Dec 15    & 0.76   & G--100 & 600 & 5 \\
2003 Dec 23    & 4.67   & G--100 & 600 & 22 \\
\noalign{\smallskip}
\multicolumn{5}{l}{\textbf{NOT 2.5\,m, ALFOSC}} \\
2003 Dec 16    & 4.12   & Grism \#7  & 600 & 24 \\
2003 Dec 17    & 6.03   & Grism \#7  & 300 & 64 \\
\noalign{\smallskip}
\multicolumn{5}{l}{\textbf{Calar Alto 3.5\,m, MOSCA}} \\
2004 Mar 01    & 0.37   & G--1000 & 300 & 4 \\
2004 Mar 02    & 2.08   & G--1000 & 450 & 15 \\
2004 Mar 03    & 3.18   & G--1000 & 450 & 14 \\
\noalign{\smallskip}\hline
\end{tabular}
\end{flushleft}
\end{table}

\subsection{\label{s-spect}Optical spectroscopy}

\paragraph{Calar Alto Observatory.} An identification spectrum of
HS\,0943+1404 was obtained on 2001 April 30 at the 2.2\,m telescope
using the Calar Alto Faint Object Spectrograph (CAFOS) equipped with
the standard SITe $2\mathrm{k}\times2\mathrm{k}$ pixel$^2$ CCD
(Table\,\ref{t-obslogspec}). The B--200 grating in conjunction with a
1.5\,\arcsec \,slit provided a spectral resolution of $\simeq10$\,\AA\
(full width at half maximum, FWHM) over the wavelength range
$\lambda\lambda3800-7000$. The spectrum was reduced online using the
\texttt{MIDAS} quicklook context available at the telescope. The
detection of strong Balmer and helium emission lines revealed the
likely CV nature of HS\,0943+1404.

CAFOS on the 2.2\,m telescope was again used in December 2002 and
December 2003 to obtain time-resolved, follow-up spectroscopy. The
G--100 grism and a slit width of 1.2\,\arcsec \,covered the wavelength
range $\lambda\lambda4300-8300$ at a spectral resolution of $\simeq
4.5$ \AA\ FWHM. The raw images were
bias-subtracted and flat-field compensated before removing the sky
contribution. The spectra of HS\,0943+1404 were optimally extracted
according to the algorithm explained in \cite{horne86-1}. To achieve a
proper wavelength calibration, a low-order polynomial was fitted to
the arc data (the {\sl rms} being less than one tenth of the
dispersion in all cases). The pixel-wavelength dependence for each
target spectrum was obtained by interpolating between the two nearest
arc spectra. The reduction and extraction were performed within
\texttt{IRAF}\footnote{\texttt{IRAF} is distributed by the National
Optical Astronomy Observatories.} and the wavelength calibration was
done in \texttt{MOLLY}.

Additional time-resolved spectroscopy was carried out in March 2004
using MOSCA on the 3.5\,m telescope. The data were obtained with the
$2\,\mathrm{k}\times4\,\mathrm{k}$ pixel$^2$ SITe CCD camera. The slit
width was fixed to 1\arcsec~and the G--1000 grism was in place. This
setup gave a wavelength range of $\lambda\lambda4200-7400$ at a
spectral resolution of $\simeq 2.6$ \AA~(FWHM). The data reduction was
carried out in the same way as described for the time-resolved CAFOS
spectroscopy.

\paragraph{Observatorio del Roque de los Muchachos.} 
The Andaluc\'\i a Faint Object Spectrograph and Camera (ALFOSC) along
with the $2048 \times 2048$ pixel$^2$ EEV chip (CCD \#8) were used at
the Nordic Optical Telescope (NOT) on La Palma to obtain time-resolved
spectroscopy of HS\,0943+1404 in December 2003. The combination of
grism \#7 (plus the second-order blocking filter WG345) and a
1\arcsec~slit provided a resolution of $\simeq 3.7$\,\AA\ (FHWM) and a
useful wavelength range of $\lambda\lambda3800-6800$. As the arc lines
projected onto more than 4 pixels (FWHM), a $1 \times 2$ binning
(dispersion direction) was applied. This provided a better
signal-to-noise ratio without significantly degrading the spectral resolution. The data reduction was carried out in the same way as explained above.

\section{Analysis: Photometry}

\subsection{Long-term variability}
From the nightly mean magnitudes presented in Table\,\ref{t-obslogphot} it is apparent that HS\,0943+1404
exhibits pronounced long-term changes in its average brightness. The
system was observed at an intermediate magnitude of $V \simeq17.6$ in
mid-December 2002, and in a faint state at $V \simeq19.0$ from
late-December 2002 until March 2003. The transition from the
intermediate to the low state hence occured in less than two weeks. By
the time of our next observations in December 2003, the system was
back to a bright state near $V \simeq 16.3$, where it apparently remained
until our last observations in March 2004. Similar long-term variability
is observed in strongly magnetic, disc-less CVs (polars) and VY\,Scl stars \citep[see e.g.][]{warner99-1}. We will hereafter follow the common terminology for these objects, calling the
bright state \textit{high state} and the faint state \textit{low
state}. 

\subsection{Short-term variability \label{s-short_term}}
In addition to the occurence of high and low states, HS\,0943+1404
displays a complex range of short-term variations which we have
analysed in detail by computing Scargle periodograms
\citep{scargle82-1} using \citeauthor{schwarzenberg-czerny89-1}'s
(\citeyear{schwarzenberg-czerny89-1}) implementation in the
\texttt{MIDAS} context \texttt{TSA}. After an initial assessment of
the periodograms, we have split the photometric data
(Table\,\ref{t-obslogphot}) into three separate subsets according to
the average brightness of the system, which we discuss below.

\subsubsection{December 2002/January 2003~--~the low state} 
\begin{figure}
\includegraphics[angle=-90,width=\columnwidth]{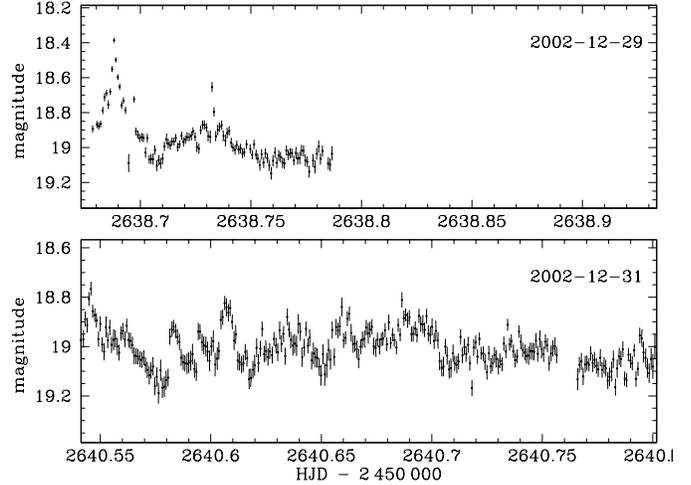}
\caption{\label{fig_lc_low} The low-state photometry obtained with the
  JKT in December 2002/January 2003 shows evidence of short-lived flares
  (top), and a low-amplitude, long-period (several hours) modulation
  superimposed on short-period variability.}
\end{figure}

\begin{figure}
\includegraphics[width=\columnwidth]{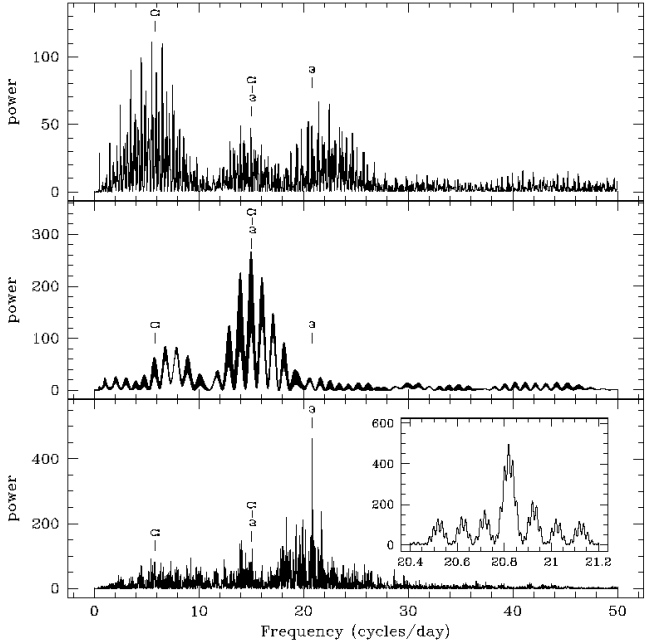}
\caption{\label{fig_scargle_phot} Scargle periodograms computed from
  three subsets of the CCD photometry listed in
  Table\,\ref{t-obslogphot}. {\em Top panel:} Periodogram for the
  low state data obtained in December 2002/January 2003. Three
  clusters of signals are evident at $\sim5\,\id$, $\sim15\,\id$, and
  $\sim22\,\id$. {\em Middle panel:} Periodogram of the December
  2003 IAC80 data and the February 2004 Tuorla data. The dominant
  signal is found at 15\,\id\ (plus one-day aliases), and a weaker signal
  at 6.8\,\id\ (plus one-day aliases). {\em Bottom panel:}
  Periodogram for the January 2004 Tuorla data and the March 2004
  campaign. The strongest signal is detected at 20.8\,\id, a second
  cluster of signals is present near 15\,\id. The inset shows a
  close-up of the 20.8\,\id\ signal. We identify the
  20.8\,\id\ frequency with the white dwarf spin $\omega$
  ($\Pspin=69.2$\,min), and the 15\,\id\ frequency with the beat between
  the white dwarf spin and the orbital motion $\omega-\Omega$, which
  implies an orbital frequency of $\Omega=5.8$\,\id\
  ($\Porb=248$\,min).}
\end{figure}

The JKT light curves of HS\,0943+1404 obtained during the low state
show short-term variability superimposed on a low-amplitude modulation
(Fig.\,\ref{fig_lc_low}). The Scargle periodogram
(Fig.\,\ref{fig_scargle_phot}, top panel) shows power in three
clusters near $\sim5\,\id$, $\sim15\,\id$, and $\sim22\,\id$. The
shortness of the individual low-state observations and their poor
sampling result in a substantial number of aliases in all three
ranges, impeding a more detailed analysis.

\subsubsection{December 2003 \& February 2004~--~a 96\,min modulation in high state} 
\begin{figure}
\includegraphics[angle=-90,width=\columnwidth]{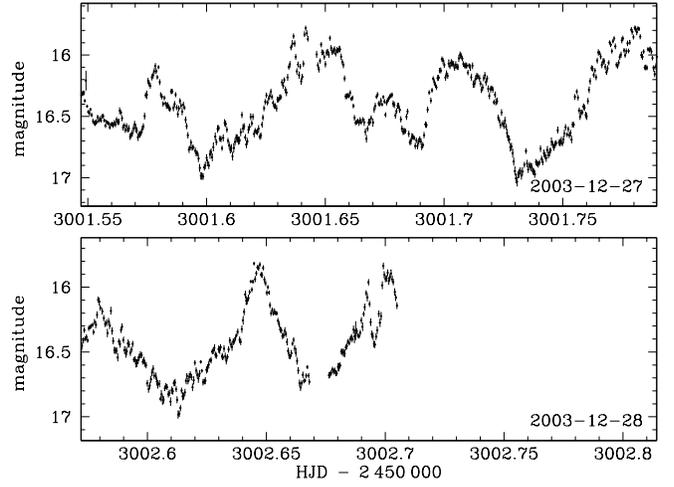}
\caption{\label{fig_lc_beat} During the two nights of IAC80
  observations, HS\,0943+1404 displayed a photometric
  variability with a period of $\simeq96$\,min, which we interpret as
  the beat period.}
\end{figure}

The light curves obtained during the December 2003 IAC80
(Fig.~\ref{fig_lc_beat}) and the February 2004 Tuorla runs display a
large-amplitude ($\simeq0.5$\,mag) modulation with a period of
$\simeq96$\,min. The Scargle periodogram of these three nights of data
(Fig.\,\ref{fig_scargle_phot}, middle panel) confirms this visual
estimate, being dominated by a signal at 15\,\id\ (plus one-day
aliases thereof). A second weaker signal is found at 6.8\,\id, also
flanked by one-day aliases. Variability on shorter time scales than
the dominant 96-min modulation is seen in the light curves, but the
periodogram contains no strong signal at higher frequencies.

\subsubsection{January \& March 2004 campaign~--~a 69\,min modulation in high state}
\begin{figure}
\includegraphics[angle=-90,width=\columnwidth]{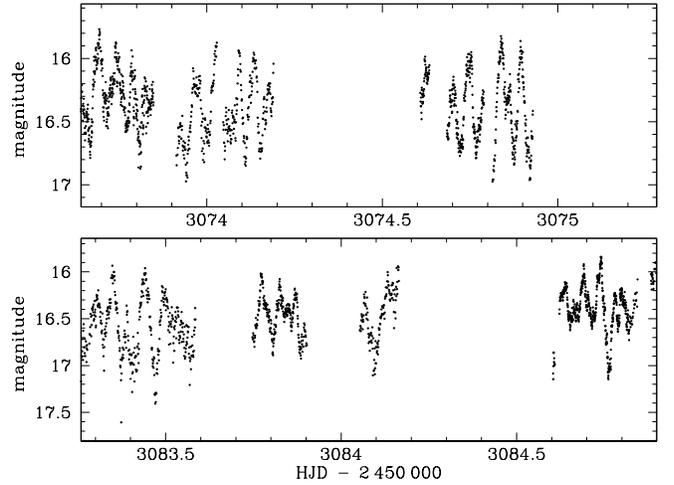}
\caption{\label{fig_lc_spin} Sample light curves obtained at FWLO,
  Hida Observatory, and Tuorla observatory during the March 2004
  campaign. The data shows a periodic modulation with a period of
  $\simeq69$\,min, which we identify with the white dwarf spin period.} 
\end{figure}

The light curves obtained in 
January 2004 at the Tuorla Observatory and during the March 2004
multi-site campaign carried out at FLWO, Hida Observatory, and Tuorla
Observatory are dominated by a coherent signal with a period of
$\simeq69$\,min and an amplitude of $0.2-0.3$\,mag
(Fig.\,\ref{fig_lc_spin}). The Scargle periodogram of this data set
(Fig.\,\ref{fig_scargle_phot}, bottom panel) is dominated by a sharp
signal at 20.8\,\id, and contains some weaker signals in the range
$14-15.2$\,\id.

\subsection{An intermediate polar interpretation}

Complex short-term variability characterized by the presence of
several coherent signals is the hallmark of IPs,
CVs containing a moderately magnetic white dwarf with a spin period
shorter than the orbital period, $\Pspin<\Porb$. The power spectra of
IPs typically show signals at the orbital frequency $\Omega$, the
white dwarf spin frequency $\omega$, and the beat (i.e. synodic)
frequency $\omega-\Omega$ (additional beat signals at $\omega+\Omega$
or $\omega-2\Omega$ have been detected in some IPs,
e.g. \citealt{warner86-2}). 

Considering the morphology of the observed light curves and the
resulting periodograms, we suggest that HS\,0943+1404 is an IP. We
identify the highly coherent, highest frequency signal detected in the
photometry of HS\,0943+1404 with the white dwarf spin, the
intermediate frequency signal as the beat between white dwarf spin and
orbital periods, and the lowest frequency signal with the orbital
period. The spin signal in the bottom panel of
Fig.\,\ref{fig_scargle_phot} shows some alias substructure due to the
long gap between the January 2004 and March 2004 data. The frequency
of the central peak is $20.8179\pm0.0003$\,\id, with the two flanking
aliases located at $20.8011\pm0.0003$\,\id\ and $20.8348\pm0.0003$\,\id,
where the errors given in brackets have been determined by fitting a
sine wave to the data. The beat signal (Fig.\,\ref{fig_scargle_phot},
middle panel) is also plagued by multiple aliases due to the
separation of the December 2003 and February 2004 runs. The strongest
signal is found at $15.0\pm0.2$\,\id, where the error in brackets is
conservatively estimated from half the FWHM of the central cluster of
aliases.

\begin{figure}
\includegraphics[angle=-90,width=\columnwidth]{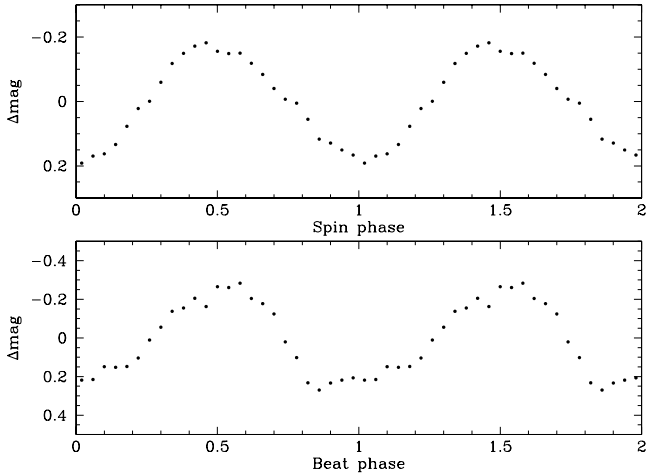}
\caption{\label{fig_folded} {\em Top panel:} The January 2004 and
  March 2004 data folded on the white dwarf spin period of
  69.2\,min. {\em Bottom panel:} The December 2003 and February
  2004 data folded on the beat period of 96\,min.}
\end{figure}

We conclude that HS\,0943+1404 is an IP, and the analysis of our
photometric data results in $\Pspin=69.171\pm0.001$\,min,
$P_\mathrm{beat}=96.0\pm3.3$\,min, and
$\Porb=247.5\pm3.3$\,min. Fig.~\ref{fig_folded} shows the spin and
beat-dominated data folded on the spin and beat period, respectively.

\section{Analysis: Spectroscopy}

\begin{figure}
\includegraphics[width=\columnwidth]{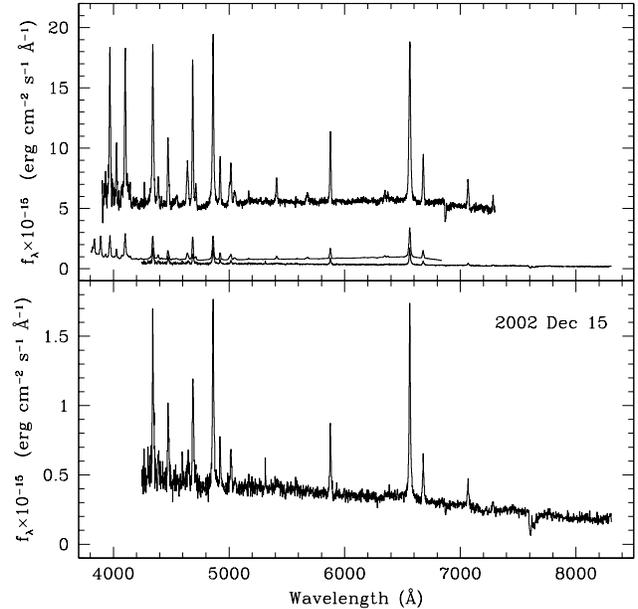}
\caption{\label{fig_idspec} {\em Top panel}: Flux calibrated average
spectra of HS\,0943+1404 taken in different epochs. From {\em bottom}
to {\em top}, 2002 December 15 (Calar Alto 2.2\,m), 2003 December
16-17 (NOT), and 2004 March 1-3 (Calar Alto 3.5\,m). These spectra
nicely illustrate the rather frequent brightness changes of
HS\,0943+0414. {\em Bottom panel}: Enlarged version of the lower spectrum shown in the top panel (2002 December). The system was in an intermediate state near $V=17.6$ mag. No spectroscopic signature of the donor star is evident in the red part of the spectrum.}
\end{figure}

\subsection{\label{s-idspec} The optical spectrum of HS\,0943+1404}
The spectrum of HS\,0943+1404 (Fig.\,\ref{fig_idspec}) is dominated by
strong, single-peaked emission lines of H\,{\sc i} and He\,{\sc
i}. \he{ii} line emission at $\lambda$4686,5412, and the
Bowen blend are also prominent, indicating the presence of a source of
high-energy photons. The relative strength of these emission lines is
reminiscent of the magnetic CVs. The continuum significantly rises
bluewards of \Ha~and atmospheric features of the secondary star are
absent in the observed spectral ranges.

\begin{figure}
\centerline{\includegraphics[width=9cm]{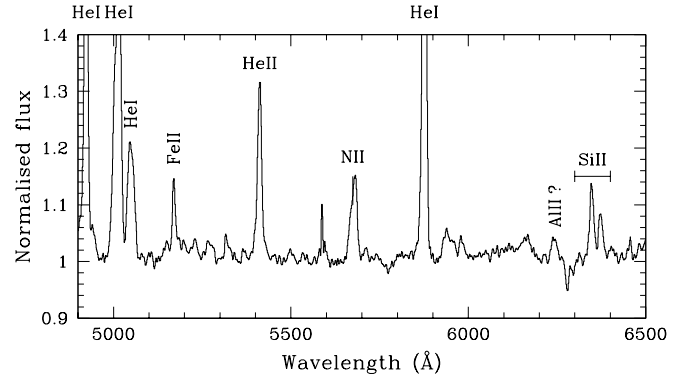}}
\caption[]{\label{f-oddlines} Average spectrum of
HS\,0943+1404 (2003 December, NOT) showing a number of
unusual emission lines in the range $\lambda\lambda4900-6500$. Note
the strong \Line{N}{II}{5680} and \Ion{Si}{II} $\lambda6347$ and
$\lambda6371$.}
\end{figure}

The most remarkable feature in the optical spectrum of HS\,0943+1404
is the strength of several emission lines that are either very weak or
absent in most other CVs, such as Fe\,{\sc ii} (e.g. $\lambda$5018,5169), \Line{N}{II}{5680}, \Ion{Si}{II} $\lambda$6347,6372, and probably \Line{Al}{II}{6243} (see the
$\lambda\lambda$4900--6500 range of the 2003 December NOT average
spectrum; Fig.~\ref{f-oddlines}). A closer inspection reveals that the
spectrum of HS\,0943+1404 shows abundant weak emission lines
throughout. In Table~\ref{table_lineparam} we present the equivalent
widths (EWs) of the most prominent lines as measured in the average
spectra corresponding to the three different epochs.

\begin{table}[t]
\caption[]{\label{table_lineparam} Line equivalent widths measured
from the average spectra.}
\begin{flushleft}
\begin{tabular}{lrclr}
\hline\noalign{\smallskip} Line & EW & ~~~~~~~~~~~~~ &Line & EW \\ &
(\AA) & & & (\AA) \\ \hline\noalign{\smallskip}
\multicolumn{5}{l}{\textbf{2002 December} (intermediate state)} \\

\Ha & $70$ & ~~~~~~~~~~~~~& \Line{He}{I}{6678} & $13$ \\ \Hb & $48$ &
~~~~~~~~~~~~~& \Line{He}{I}{5876}& $18$ \\ \Hg & $33$ & ~~~~~~~~~~~~~&
\Line{He}{I}{4922} & $7$ \\ \Line{He}{II}{4686}& $25$ & ~~~~~~~~~~~~~&
\Line{He}{I}{4472} & $11$ \\ \Line{He}{II}{5412}& $4$ & ~~~~~~~~~~~~~&
\Line{He}{I}{4388} & $3$ \\ Bowen blend & $6$ & ~~~~~~~~~~~~~&
\Line{N}{II}{5680} & $2$ \\ \Line{He}{I}{7065} & $10$ & ~~~~~~~~~~~~~&
& \\
\noalign{\smallskip} \multicolumn{5}{l}{\textbf{2003 December} (high state)} \\
\Ha & $53$ & ~~~~~~~~~~~~~& \Line{He}{I}{7281} & $5$ \\ \Hb & $49$ &
~~~~~~~~~~~~~& \Line{He}{I}{7065} & $9$ \\ \Hg & $44$ & ~~~~~~~~~~~~~&
\Line{He}{I}{6678} & $12$ \\ \Hd & $42$ & ~~~~~~~~~~~~~&
\Line{He}{I}{5876} & $16$\\ H$\varepsilon$ & $32$ & ~~~~~~~~~~~~~&
\Line{He}{I}{4922} & $10$\\ H8 & $22$ & ~~~~~~~~~~~~~&
\Line{He}{I}{4472} & $16$ \\ H9 & $13$ & ~~~~~~~~~~~~~&
\Line{He}{I}{4388} & $7$ \\ \Line{He}{II}{4686}& $35$ & ~~~~~~~~~~~~~&
\Line{He}{I}{4026} & $11$ \\ \Line{He}{II}{5412}& $6$ & ~~~~~~~~~~~~~&
\Line{N}{II}{5680} & $4$ \\ Bowen blend & $16$ & ~~~~~~~~~~~~~& & \\
\noalign{\smallskip} \multicolumn{5}{l}{\textbf{2004 March} (high state)} \\
\Ha & $52$ & ~~~~~~~~~~~~~& \Line{He}{I}{6678} & $13$ \\ \Hb & $51$ &
~~~~~~~~~~~~~& \Line{He}{I}{5876} & $14$\\ \Line{He}{II}{4686}& $31$ &
~~~~~~~~~~~~~& \Line{He}{I}{4922} & $10$\\ \Line{He}{II}{5412}& $8$ &
~~~~~~~~~~~~~& \Line{He}{I}{4472} & $19$ \\ Bowen blend & $17$ &
~~~~~~~~~~~~~& \Line{He}{I}{4388} & $9$ \\ \Line{He}{I}{7065} & $9$ &
~~~~~~~~~~~~~& \Line{N}{II}{5680} & $5$ \\ \noalign{\smallskip}\hline
\end{tabular}
\end{flushleft}
\end{table}

\subsection{Radial velocity variations}

\begin{figure}
\centering \includegraphics[width=9cm]{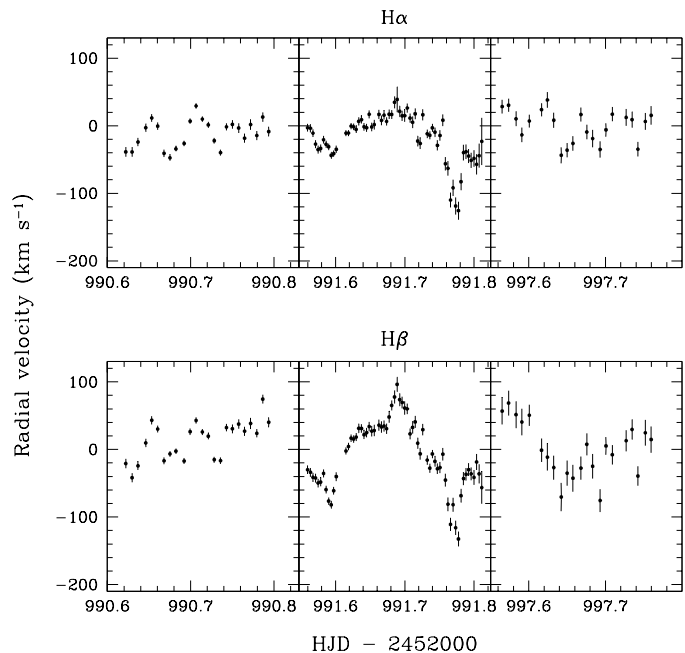}
\caption{\label{fig_vel} \Ha~({\em top}) and \Hb~({\em bottom}) radial
velocity curves. From {\em left} to {\em right}: 2003 December 16, 17,
and 23.}
\end{figure}

In order to search for orbital variability, we measured the radial
velocities of the \Ha, \Hb, \Line{He}{I}{5876} and \Line{He}{II}{4686}
emission lines. The spectra were first re-binned to a
constant velocity scale centred at the rest wavelength of each
line and normalised. Radial velocities were measured by using three different
methods: (i) correlation with a single Gaussian template, (ii) the
double Gaussian technique of \cite{schneider+young80-2}, and (iii)
calculation of the median velocity (the point leaving the same area on
both sides of the normalised line profile). The latter method proved to be
the most efficient, producing the cleanest radial velocity curves. The
resulting curves for \Ha~and \Hb~are shown in Fig.~\ref{fig_vel}.

The morphology of the radial velocity curves seems to significantly
change from night to night. On 2003 December 16, a modulation at $\sim
70$ min dominates, whereas a longer time scale modulation is clearly seen on 2003
December 17. On this night, a dip towards blue velocities at HJD $\simeq
2452991.77$ is observed. A shallower excursion to the blue also occurs
at HJD $\simeq 2452991.59$. The separation between them was measured by
fitting Gaussians to the dips, obtaining a value of $\simeq 0.19$\,d. In
addition, the \Hb~curve displays a red spike centred at $\simeq
2452991.69$, almost just in the middle of the blue dips. A sine fit to
the \Hb\ radial velocity curve (after masking the dips) yields a period
of $\simeq 0.19$\,d and a semi-amplitude of $\sim 45$\,\kms.

\begin{figure}
\centering \includegraphics[width=9cm]{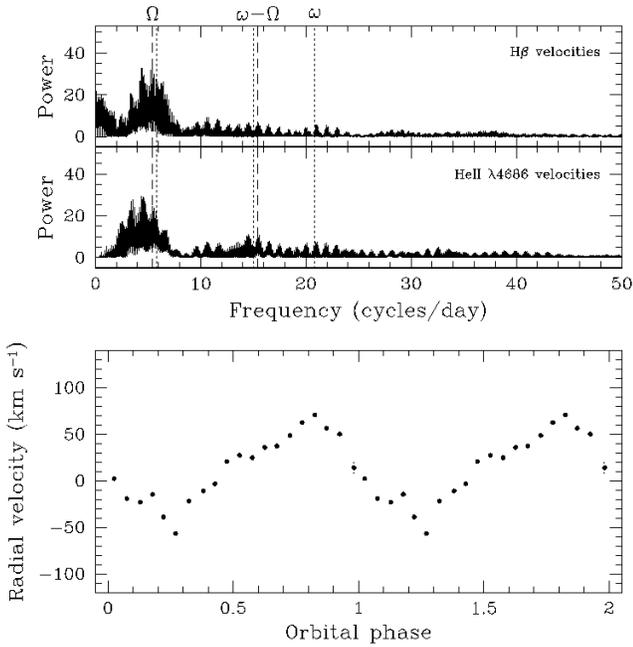}
\caption{\label{fig_hbscargle} {\em Top}: Scargle periodograms of the
\Hb~and \Line{He}{II}{4686} radial velocity curves. $\Omega$ is the
orbital frequency, $\omega$ the white dwarf spin, and $\omega-\Omega$
the corresponding synodic frequency. Dashed (dotted) vertical
lines correspond to spectroscopic (photometric) results. {\em
Bottom}: \Hb~radial velocity curve folded on 265.2 min after
averaging the data into 20 phase bins. The time of red-to-blue
crossing is $T_0=2452623.78160$ (HJD). The orbital cycle has been
plotted twice for continuity.}
\end{figure}

In order to search for periodicities we computed Scargle periodograms
\citep{scargle82-1} for all the \Ha, \Hb, \Line{He}{I}{5876}, and
\Line{He}{II}{4686} radial velocity curves. The periodograms for all
the lines are very similar, and we present in Fig.~\ref{fig_hbscargle}
only the periodograms for \Hb\ and \Line{He}{II}{4686}. Both
periodograms are dominated by clusters of aliases near $\simeq 3.4$,
4.4, and 5.4\,\id. The poor sampling of our radial velocities spread
out over 15 months causes severe fine-structure in the main alias
clusters, and impedes an unequivocal identification of the frequency
inherent to the system. Formally, the strongest signal in the \Hb\
radial velocity variation is found at $5.430 \pm 0.002$\,\id\ (the
quoted error is a conservative estimate from the width of the peak in
the periodogram), which corresponds to a period of $P=0.18416 \pm
0.00007$ d ($= 265.2 \pm 0.1$ min). In the case of \Line{He}{II}{4686}, a value of $4.412 \pm 0.005$\,\id~($P=0.2267 \pm 0.0003 \,\mathrm{d} = 325.8 \pm
0.4$\,min) is found. The presence of power around 250\,min in both
the low-state photometry (Fig.\,\ref{fig_scargle_phot}, top panel) and
in the radial velocity variations (Fig.\,\ref{fig_hbscargle}) strongly
suggests the detection of a ``clock'' in HS\,0943+1404. Given that this
is the longest period detected in all our data we believe that it
represents the orbital motion of the binary.

The \Hb\ radial velocities folded on 265.2\,min, the strongest signal
in the \Hb\ periodogram, displays a relatively low-amplitude
modulation superimposed by a short ($\sim0.15$ orbital phase) negative
excursion, which is mostly related to the two ``dips'' seen in the
December 17 NOT data.

\subsection{$V/R$ ratios and equivalent width curves}

\begin{figure}
\centering \includegraphics[width=9cm]{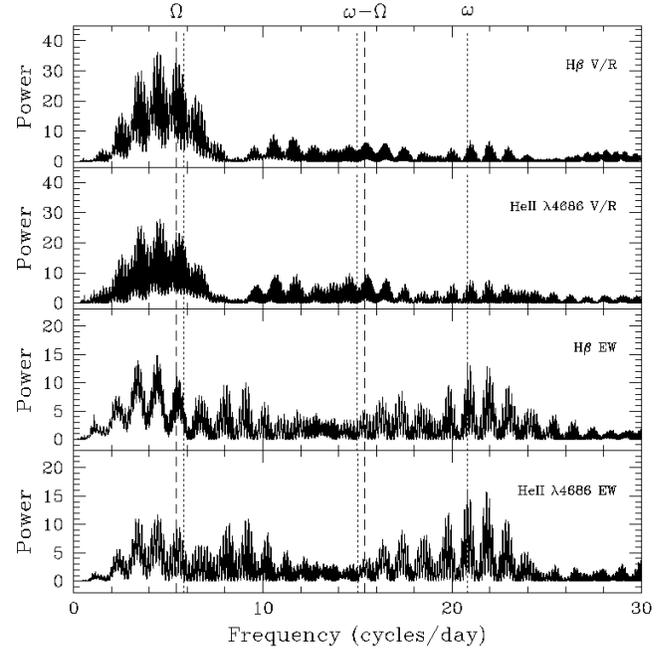}
\caption{\label{fig_vr_ew_scargle} Scargle periodograms of the \Hb~and
\Line{He}{II}{4686} $V/R$ and EW curves. Dashed (dotted) vertical
lines correspond to spectroscopic (photometric) results.}
\end{figure}

Our second approach at searching for periodicities involved the
analysis of the \Ha, \Hb, \Line{He}{I}{5876} and \Line{He}{II}{4686}
$V/R$-ratio and EW curves. The $V/R$ ratio is computed by dividing the EW of the blue half of the line profile (up to zero velocity, i.e. the $V$ part) by that of the red half (redwards from zero velocity; the $R$ part) Only the Scargle periodograms for \Hb\ and
\Line{He}{II}{4686} are shown in Fig.~\ref{fig_vr_ew_scargle}. The
picture in the low frequency regime does not shed new light on the
orbital period discussion, as it contains the same three clusters of
aliases around 3.4, 4.4, and 5.4\,\id. However, the EW periodograms
contain a strong peak at the photometric spin frequency. The strongest peak in both the \Hb\ and \Line{He}{II}{4686} EW periodograms is centred at 20.82\,\id, exactly at the spin frequency.

\section{Discussion}

\subsection{The intermediate polar nature of HS\,0943+1404}

The combined photometric and spectroscopic results clearly qualify
HS\,0943+1404 as an asynchronous magnetic CV, an IP. The system has a likely
orbital period of $\Porb\simeq250$\,min and a white dwarf spin period of $\Pspin = 69.171 \pm 0.001$\,s. The radial velocity analysis of our spectroscopy suggests $\Porb=265.2$\,min, while the more extensive photometry suggests $\Porb=247.5$\,min from the detection of the beat and spin periods.  

The combination of orbital and spin periods found in HS\,0943+1404 is
rather unusual, as it yields a large $\Pspin/\Porb\simeq 0.3$ and a long
orbital period. The only other known IPs with even larger
$\Pspin/\Porb$ are EX\,Hya \citep{vogtetal80-2}, V1025\,Cen
\citep{buckleyetal98-1}, and DW\,Cnc \citep{rodriguez-giletal04-1},
all three are short-period systems below the period gap. The majority
of IPs are concentrated near $\Pspin/P_\mathrm{orb} \simeq 0.1$
\citep[see e.g.][]{barrettetal88-1, king+lasota91-1,
warner+wickramasinghe91-1}, but \cite{king+wynn99-1} showed that
spin-orbit equilibria exist with $\Pspin/P_\mathrm{orb} > 0.1$,
explaining cases such as EX\,Hya. Most recently, \cite{nortonetal04-1}
demonstrated that a large range of permitted spin-orbit equilibria
exists in the ($\Pspin/P_\mathrm{orb}$, $\mu_1$) parameter space, with
$\mu_1$ the magnetic moment of the white dwarf primary star. We
used their Figure\,2 to estimate a magnetic moment of $\mu_1 \sim
10^{34}\,\mathrm{G\,cm^3}$ for HS\,0943+1404. This value is in the
typical range found for polars, and assuming an average
white dwarf ($\Mwd\simeq0.65\,\Msun$, $\Rwd\simeq10^8$\,cm) would
result in a magnetic dipole field strength of $\sim20$\,MG. Hence, it appears entirely possible that HS\,0943+1404 is a true ``intermediate''
polar, that will synchronise once its orbital period and mass
transfer rate decrease sufficiently~--~in fact, for the parameters determined here, HS\,0943+1404 lies above the synchronisation line in Figure 4 of \cite{nortonetal04-1}. 

Another property that HS\,0943+1404 shares with polars is the
occurence of deep ( $\simeq 3$\,mag) low states, such as the one
observed in January 2003 (Table\,\ref{t-obslogphot}). Such low states
are characteristic of polars, only one long-period IP is known to have
entered a deep low state \citep[see e.g.][]{warner99-1}.

\subsection{Spin-beat period switching}
Our photometric data show that HS\,0943+1404 switches between states
which are either dominated by variability at the white dwarf spin
period, or by variability at the beat period.  Changes in the relative
power of the spin signal with respect to the side-band signals have
been observed in other IPs, such as TX Col \citep{buckley+tuohy89-1,
nortonetal97-1}, and in V1062\,Tau, where at a given time only the spin
or the beat period is detected \citep{lipkinetal04-1}. In V1062\,Tau,
\citet{lipkinetal04-1} suggested that the switching between spin- and
beat-dominated variability is triggered by changes in the system
brightness which they relate to changes in the mass transfer rate.

A dominating signal at the beat period has been interpreted as a sign
of disc-less accretion (e.g. \citealt{buckleyetal95-1};
\citealt{hellier02-1}, and references therein). Spin-dominated
variability is generally explained by disc-fed accretion. The variations are thought to arise from reprocessing of X-rays at either the inner edge or the
outer rim of the magnetically-truncated disc \citep{warner86-2}, or
from aspect variations of the accretion funnel feeding material from
the inner disc edge onto the white dwarf \citep[see e.g.][]{hellieretal87-1}.

The alternating dominance of the spin and beat signals in the light
curves of HS\,0943+1404 clearly implies changes in the accretion mode. However, in this system, both states have been observed at similar mean brightness levels (Table\,\ref{t-obslogphot}, Fig.\,\ref{fig_lc_beat}, Fig.\,\ref{fig_lc_spin}). The fact that the spectrum of the secondary star is not detected even when the system is faint (although no spectrum in low state is available; see Fig.~\ref{fig_idspec}) indicates that the accretion disc must dominate the observed continuum. Hence, a predominant signal at the beat period cannot be unambiguously linked to disc-less accretion. All of this suggests that changes in the accretion mode are not solely due to variations in the mass transfer rate, but can also be triggered by other mechanisms within the binary system \citep{nortonetal97-1}.

\subsection{An elevated nitrogen abundance}
The optical spectrum of HS\,0943+1404 is similar to that of the IP
1RXS\,J062518.2+733433 (= HS\,0618+7336;
\citealt{araujo-betancoretal03-2}). Both systems show unusually strong
emission of \Line{N}{II}{5680} compared to most other CVs. We
interpret this as evidence for an enhanced nitrogen abundance in these two
IPs, which is usually associated with the existence of CNO-processed
material in the envelope of the donor star. Two possible scenarios can
explain the origin of the CNO-processed material in the envelope of
the donor: either the companion has accreted material from the shell
associated with a nova explosion on the white dwarf, or the secondary
star is evolved (and the white dwarf is accreting material
from the exposed CNO core whose outer layers were stripped off during
a thermal time-scale mass-transfer phase, see
\citealt{schenkeretal02-1,podsiadlowskietal03-1}). Based on a HST
FUV survey of cataclysmic variables, \cite{gaensickeetal03-1} have
recently shown that 10\,\%--15\,\% of all CVs display enhanced
nitrogen abundances. This is in agreement with the predictions by
\citet{schenkeretal02-1} and \citet{podsiadlowskietal03-1} that a non-negligible
fraction of all CVs should have evolved through a phase of thermal
time scale mass transfer.

\section{Conclusions}

The results presented in this paper are summarised as follows:

\begin{enumerate}
\item Based on our photometric and spectroscopic observations, we have
shown that HS\,0943+1404 is a new intermediate polar CV with an
orbital period of $\Porb\simeq250$\,min and a white dwarf spin period
of $\Pspin = 69.171 \pm 0.001$\,s.  An additional confirmation of the
spin period, e.g. through the detection of a coherent variablity in
X-rays or polarised light would be desirable.
\item Based on \citeauthor{nortonetal04-1}'s (\citeyear{nortonetal04-1})
calculations, we estimate the magnetic moment of the white dwarf in
HS\,0943+1404 to be $\mu_1\sim10^{34}\,\mathrm{G\,cm^{3}}$, which is
in the range of polars. Long-term photometric monitoring of the system
revealed the occurrence of deep low states, during which the
brightness drops by $\sim 3$\,mag, also typical of polars. It appears
likely that HS\,0943+1404 is a true ``intermediate'' polar on its way
to synchronization.
\item The optical spectrum of HS\,0943+1404 contains a number of
unusual emission lines, most noticeably \Line{N}{II}{5680}, which 
suggests the presence of CNO-enriched material in the envelope of the
secondary star. Ultraviolet spectroscopy of HS\,0943+1404 would be
useful to assess the carbon and nitrogen abundances from the
\Line{N}{V}{1240} and \Line{C}{IV}{1550} resonance lines.
\end{enumerate}

\begin{acknowledgements}
We wish to thank the anonymous referee for his/her contribution to the paper. We are grateful to Sergio Fern\'andez for carrying out some of the IAC80 telescope observations. PRG and BTG thank PPARC for support through a PDRA and an AF, respectively. The HQS was supported by the Deutsche Forschungsgemeinschaft through grants Re\,353/11 and Re\,353/22. This work was supported in part by NASA grant NAG5-9930. The use of the \texttt{MOLLY} package developed and maintained by Tom Marsh is acknowledged.  
\end{acknowledgements}

\bibliographystyle{aa} 
\bibliography{aamnem99,aabib} 
\end{document}